\documentclass[twocolumn,amsmath,amssymb,superscriptaddress,aps,prb]{revtex4-1}

\usepackage{lmodern}
\usepackage[utf8]{inputenc}
\usepackage[english]{babel}
\usepackage{graphicx}
\usepackage{color}
\usepackage{hyperref}
\usepackage[all]{hypcap}    

\begin{document}

\title{The origin of the core-level binding energy shifts in nanoclusters}

\author{Alexey A. Tal}
\email{aleta@ifm.liu.se}
\affiliation{Department of Physics, Chemistry and Biology (IFM), Linköping University, SE-581 83, Linköping, Sweden}
\affiliation{Materials Modeling and Development Laboratory, National University of Science and Technology 'MISIS', 119049, Moscow, Russia}

\author{Weine Olovsson}
\affiliation{Department of Physics, Chemistry and Biology (IFM), Linköping University, SE-581 83, Linköping, Sweden}%

\author{Igor A. Abrikosov}
\affiliation{Department of Physics, Chemistry and Biology (IFM), Linköping University, SE-581 83, Linköping, Sweden}%

\begin{abstract}
We investigate the shifts of the core-level binding energies in small gold nanoclusters by using {\it ab initio} density functional theory calculations. The shift of the 4$f$ states is calculated for magic number nanoclusters in a wide range of sizes and morphologies. We find a non-monotonous behavior of the core-level shift in nanoclusters depending on the size. We demonstrate that there are three main contributions to the Au 4$f$ shifts, which depend sensitively on the interatomic distances, coordination and quantum confinement. They are identified and explained by the change of the on-site electrostatic potential.
\end{abstract}

\maketitle

\section{\label{sec:Introduction}{Introduction}}
Nanoclusters with the size of nanometers demonstrate fascinating reactive, optical, electronic and magnetic properties, which are not observed in their bulk counterparts\cite{Baletto2005}. That behavior is determined by many factors such as high surface-to-bulk ratio, electronic shell closing \cite{Kappes1982,Ekardt1984,Knight1984}, geometric shell closing \cite{Kappes1982} and quantum confinement\cite{Efros1996}. Spectroscopic measurements, especially such precise as x-ray photoelectron spectroscopy (XPS), have been widely used for the characterization of small nanoclusters\cite{Ferrando2008}. The properties of nanoclusters make them extremely interesting for catalysis applications, where reactions can be altered by small changes in structure or size. The high ratio of surface-to-bulk atoms in the clusters drastically increases their efficiency. In the recent paper by X. Ma et al. \cite{Ma2017} the local coordination of gold atoms in small nanoclusters has been suggested as a parameter for catalytic activity prediction. Moreover, in the work of W.E. Kaden {\it et al.}\cite{Kaden2009} it was shown that shifts of the binding energies of core-electrons strongly correlate with the catalytic activity of the nanoclusters. Thus, understanding of the origin of core-level binding energy shifts (CLS) in nanoclusters of different size is important for their characterization, with a potential for designing nanoparticles with improved performance.

Morphologies of small nanoclusters are very different from their corresponding bulk structures\cite{Baletto2005}. Thermodynamically favorable morphology of a nanocluster is determined by the competition of surface energy and internal stress. The analysis of the thermodynamics of the gold nanoclusters performed by F. Baletto {\it et al.}\cite{Baletto2002} showed that icosahedral and decahedral structures are the most favorable for small nanoclusters. However, due to kinetics in the growth process, cubic and octahedral clusters are present as well\cite{Chen2005}. In this work we consider CLS in clusters with magic number of atoms with cubic, icosahedral, decahedral and octahedral morphologies. During the growth process kinetics might not allow nanoclusters to transition into favorable morphology\cite{tal2015}. Moreover, a substrate where nanoclusters are collected affects the morphology and may induce a shift of the core states due to charge transfer\cite{Lykhach2016a}. All these effects significantly complicate the analysis of nanoclusters. The use of theoretical modeling allows to distinguish trends and illuminate on the origin of positions of the core states in nanoclusters. Thus, we investigate core-level shifts of unsupported neutral nanoclusters with ideal structures, but it is worth emphasizing that clusters of larger size should not be significantly affected by the substrate.
The great interest of gold nanoclusters in applications for catalysis motivated our choice of the material\cite{Coquet2008,Tyo2015,Meyer2004,Zhang2014}. Moreover, gold has the largest surface core-level shift of all noble metals \cite{Citrin1978}, which makes it a good candidate for studying the behavior of the core-levels and easier to distinguish trends.

Several experiments have demonstrated that 4$f$ levels shift towards higher binding energies in Au nanoclusters with the decrease of the nanocluster size\cite{Howard2002,Wertheim1983,Dalacu2001}. Conventionally, these shifts are believed to be due to the final state relaxation induced by charging\cite{Wertheim1983}. Besides the shift, a broadening of the 4$f$ peak was observed in all experiments, explained by the effect of the electrostatic potential. The negative surface core-level shifts observed in experiments are to a large extent an initial state effect, explained by the valence band narrowing of the less coordinated surface atoms\cite{Citrin1978}. This causes a charge redistribution from 6$s$ to 5$d$ and hence an increased charge density and screening of the core-hole\cite{Citrin1983b, Andersen1994}. However, the structural effect on the CLS has not been fully understood. These aspects motivate the present work. In particular, we systematically study the structural effects on the shifts of the core-levels and the relation between initial and final states. Moreover, we investigate the evolution of the core-state energies from an individual atom to bulk systems through atomic clusters. Although, the charge induced shift can be significant in the spectra, we focus our attention on structural effects.

First-principles calculations of core-level shifts have proven to be a very accurate and useful tool for understanding of the XPS spectra and behavior of the core-level in general\cite{Marten2005,Gronbeck2012}. Here we demonstrate that our calculations for the surface core-level shifts in Au accurately reproduce the experimental values \cite{Citrin1978}. Based on this, we study how energy levels of the core-states depend on the morphology and size of the nanoclusters and how these levels change for different atoms in a nanocluster. Furthermore, the change of the core-levels is analyzed in the sequence atom-cluster-bulk.

This paper is organized as follows. In Section \ref{sec:CompDet} the methods and details of the calculations are provided. The results are discussed in section \ref{sec:Results}. Subsection \ref{sec:Effects} describes the effects of the strain, coordination and size on the shift of the core-states. Subsection \ref{sec:Cluster} is dedicated to CLS in icosahedral, decahedral and octahedral nanoclusters. Finally, Section \ref{sec:Conclusion} presents our conclusions.

\section{\label{sec:CompDet}{Computational details}}
We performed {\it ab initio} density functional theory (DFT) calculations of the core-level binding energy shifts\cite{Olovsson2010} within the complete screening picture, in doing so we include both initial (the shift of the on-site electrostatic potential for an atom in different environments) and final (core-hole screening by conduction electrons) state effects. The electron wave functions were treated within the projector augmented wave (PAW) method\cite{Blochl1994}. The plane waves cut off energy was 250~eV. The value of the cut off was determined from convergence of core-level shifts in bulk and nanoclusters. The integration of the Brillouin zone was performed in $\Gamma$ point for clusters larger than 50 atoms, and with denser k-point grid for small nanoclusters. The DFT calculations were performed with GGA exchange-correlation functional in PBE\cite{Perdew1996} form as implemented in the Vienna ab-initio simulation package, {\tt VASP}\cite{Kresse1999}. For the {\it initial state} approximation CLS calculations, the Kohn-Sham equation is solved inside the PAW sphere for core electrons, after self-consistency with frozen core electrons has been attained\cite{Kohler2004}. In calculations with the core-hole, we assume that the the 0hole at the ionized atom effectively acts as an extra proton. This assumption allows us to substitute the ionized atom of atomic number Z with the next element in the Periodic Table. This approximation is also called equivalent core or (Z+1)-approximation\cite{Johansson1980,Marten2005,Olovsson2010}.
In calculations CLS can be determined from the difference between ionization energies, which is the difference between total energy in the ground state and total energy of the core-ionized state. Thus, ionization energy can be defined in the form of a generalized thermodynamic chemical potential (GTCP)\cite{Johansson1980},
\begin{equation}
 \mu = \frac{E^{ion}-E^{gs}}{1/N},
\end{equation}
where $\mu$ is the GTCP and N is the number of ionized atoms in the supercell ($N = 1$ in our calculations). The energies $E^{gs}$ and $E^{ion}$ are the total energies of the system in the ground state and ionized states respectively. The shifts can be calculated from:
\begin{equation}
 E_{CLS}=\Delta\mu_i = \mu_i-\mu_i^{Ref},
\end{equation}
where $i$ is the core-level in the atom of the study. $\mu_i^{Ref}$ corresponds to a reference system. In this study, we have chosen pure bulk fcc Au as a reference system. It is important to notice that the shifts are calculated relative to the Fermi level. The {\it complete screening} CLS approximation is known to be reliable and to reproduce experiments well\cite{Marten2005,Cavallin2012}.

\section{\label{sec:Results}{Results and discussion}}
For reasons of clarity and comparison to other theoretical calculations as well as to experiment, we will forthwith discuss the 4$f$ states. The core-states are very sensitive to the change of the local environment of an atom and, as will be shown below, the three main contributions to the shifts are: number of nearest neighbors (coordination), confinement and the lattice parameter. We have found that the shifts correlate with the behavior of the $d$-band and can be explained by the change of the  on-site electrostatic potential as shown below.

\subsection{\label{sec:Effects}{Contributions affecting shifts of the core states}}

\begin{figure}[h!]
\includegraphics[width=0.5\textwidth]{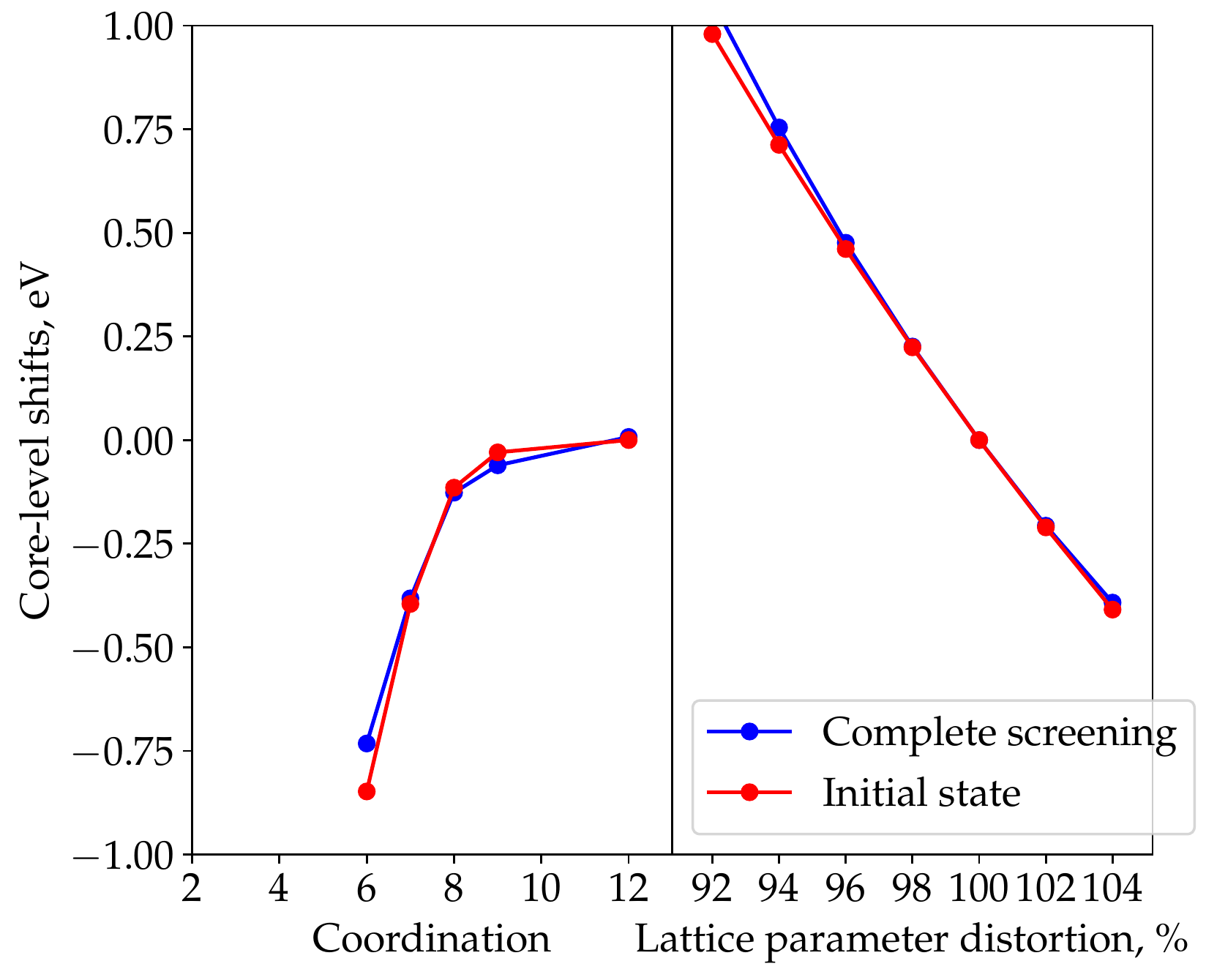}
\label{fig:strain}
\caption{ a) Au 4$f$ core-level shifts of atoms with different coordination (number of nearest neighbors). Reduced coordinations are obtained by cutting (124), (112), (100), (111) surfaces. b) Au 4$f$ core-level shifts of atoms in bulk structure with distorted lattice parameter. Distortion is denoted in percent of the perfect bulk lattice parameter.}
\end{figure}

\paragraph{Coordination effects}
Fig.~\ref{fig:strain}a shows how the position of the 4$f$ core-state depends on the coordination of the atom, where a coordination of 12 corresponds to an atom in the bulk and has zero CLS. These calculations were performed for nanoclusters with fcc structure and with the bulk lattice parameter. Atoms from surfaces (124), (112), (100), (111) were chosen as undercoordinated atoms. The largest shift was found for the most undercoordinated atoms. The shift decreases for atoms with larger coordination and saturates for coordination of 10. The shift with and without final state effect are very similar. This means that the effects of the core-hole screening are not so significant.
\paragraph{Strain effects}
As shown in Fig.~\ref{fig:strain}b 4$f$ states are very sensitive to the strain. Under uniform compression 4f state shifts toward higher binding energies, and in the opposite direction for uniformly stretched lattice parameter. The change of the lattice parameter by 2\% results in the shift of 0.25 eV. Once again, the shifts calculated within initial and complete screening are very similar.

\begin{figure}[ht!]
\includegraphics[width=0.4\textwidth]{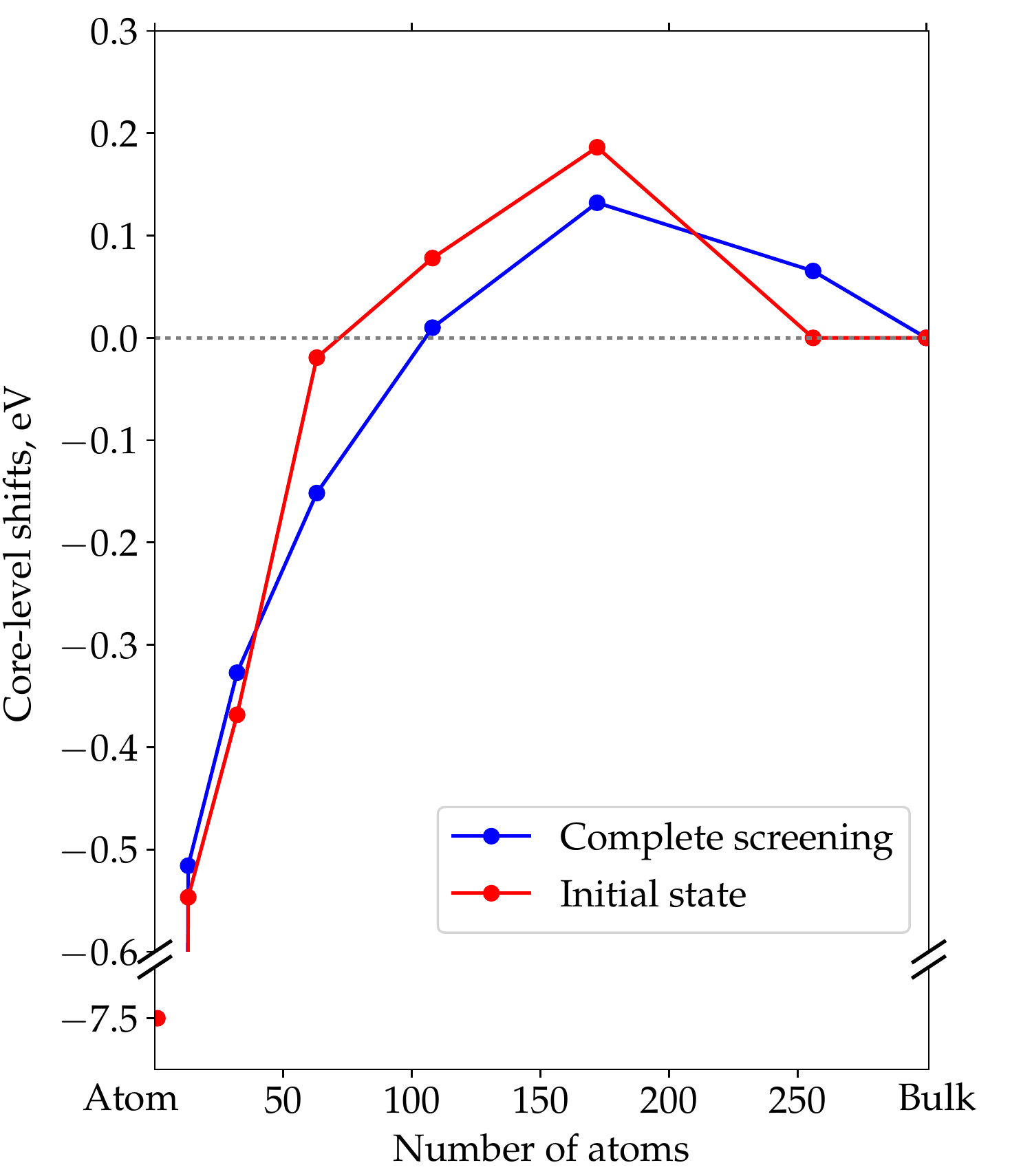}
\label{fig:conf}
\caption{Au 4$f$ core-level shift as a function of the size of a nanocluster within initial state and complete screening approximations. The structures are fcc unrelaxed clusters, where position of the 4$f$ states is calculated for central atom.}
\end{figure}

\paragraph{Size effect}
Another parameter that affects CLS in nanoclusters is the size effect or confinement effect. In Fig~\ref{fig:conf} the results for the calculated 4$f$ state are shown. The shifts were calculated for the central atom in cubic with ideal (unrelaxed) fcc structure containing clusters from 13 to 256 atoms. Starting from the smallest nanocluster with 13 atoms, 4$f$ states are shifted towards smaller binding energies. Then for larger clusters this shift becomes smaller in magnitude and approaches the position of 4$f$ state. It is important to emphasize that the shift non-monotonically depends on the size and even changes the sign at around 100 atoms.

\begin{figure}[ht!]
\includegraphics[width=0.5\textwidth]{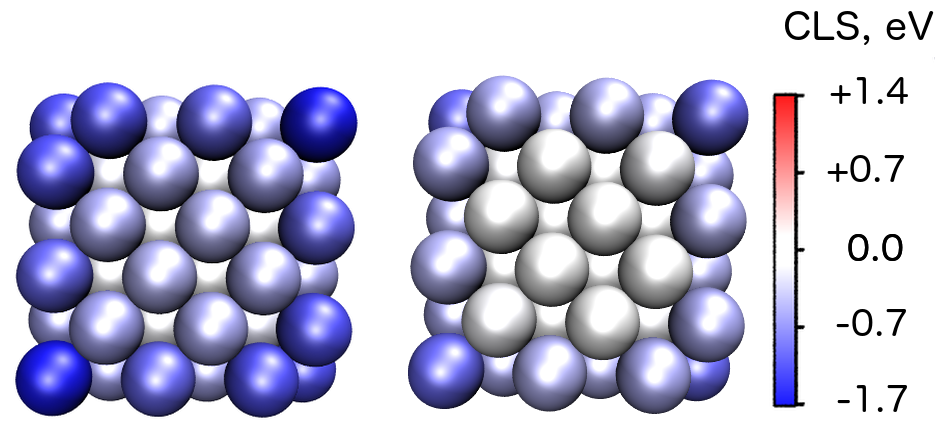}
\label{fig:108}
\caption{Shifts of 4$f$ states in 108 cubic Au nanocluster with bulk lattice parameter on (100) facet (left) and its cross section (right).   Colors of atom correspond to the value of the shift. }
\end{figure}

\begin{figure}[ht!]
\includegraphics[width=0.5\textwidth]{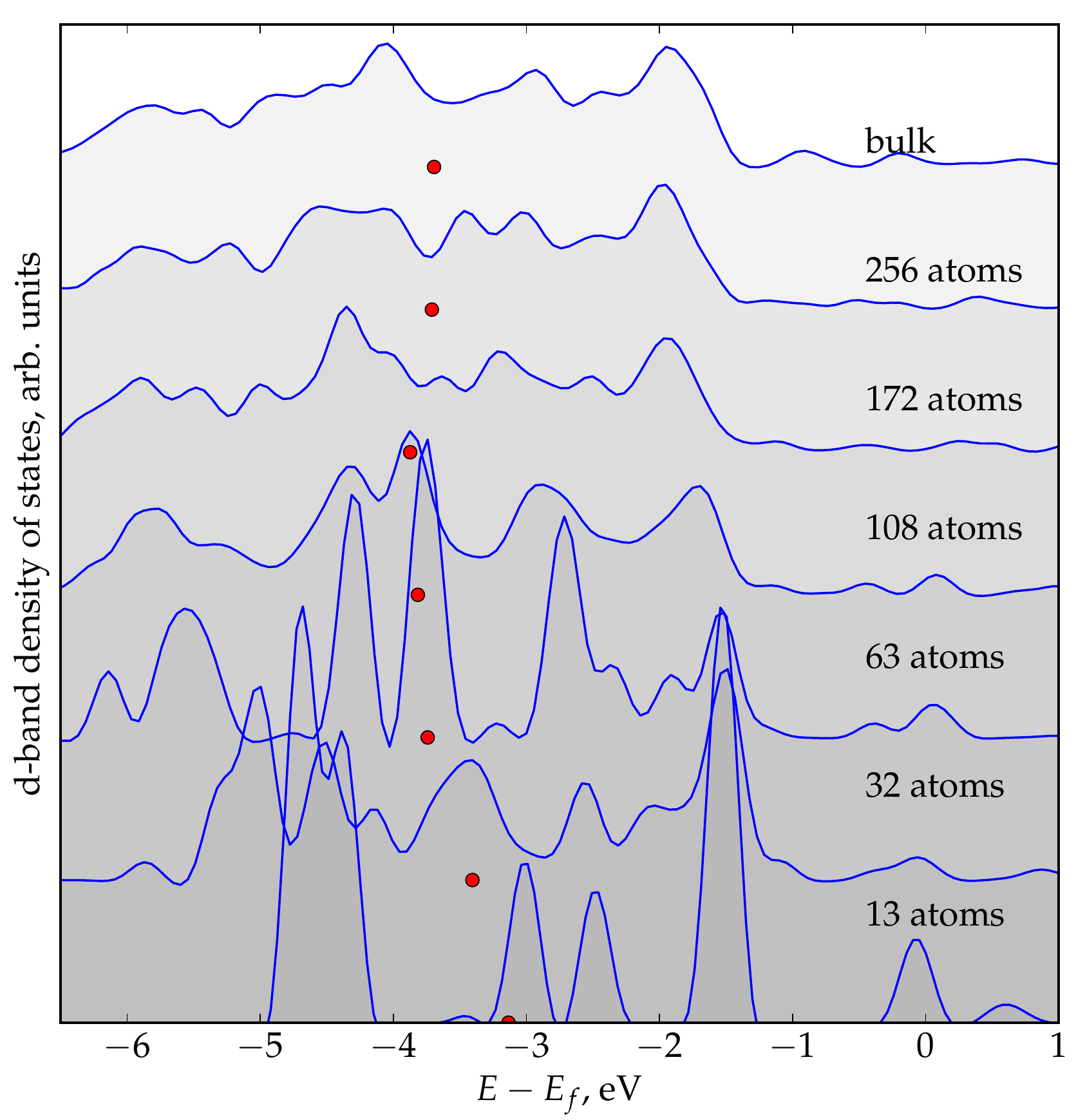}
\label{fig:dos_conf}
\caption{Density of states in $d$-band of the central atom in a nanoclusters with different number of atoms, where red dots denote the center of $d$-band.}
\end{figure}

\begin{figure}[ht!]
\includegraphics[width=0.50\textwidth]{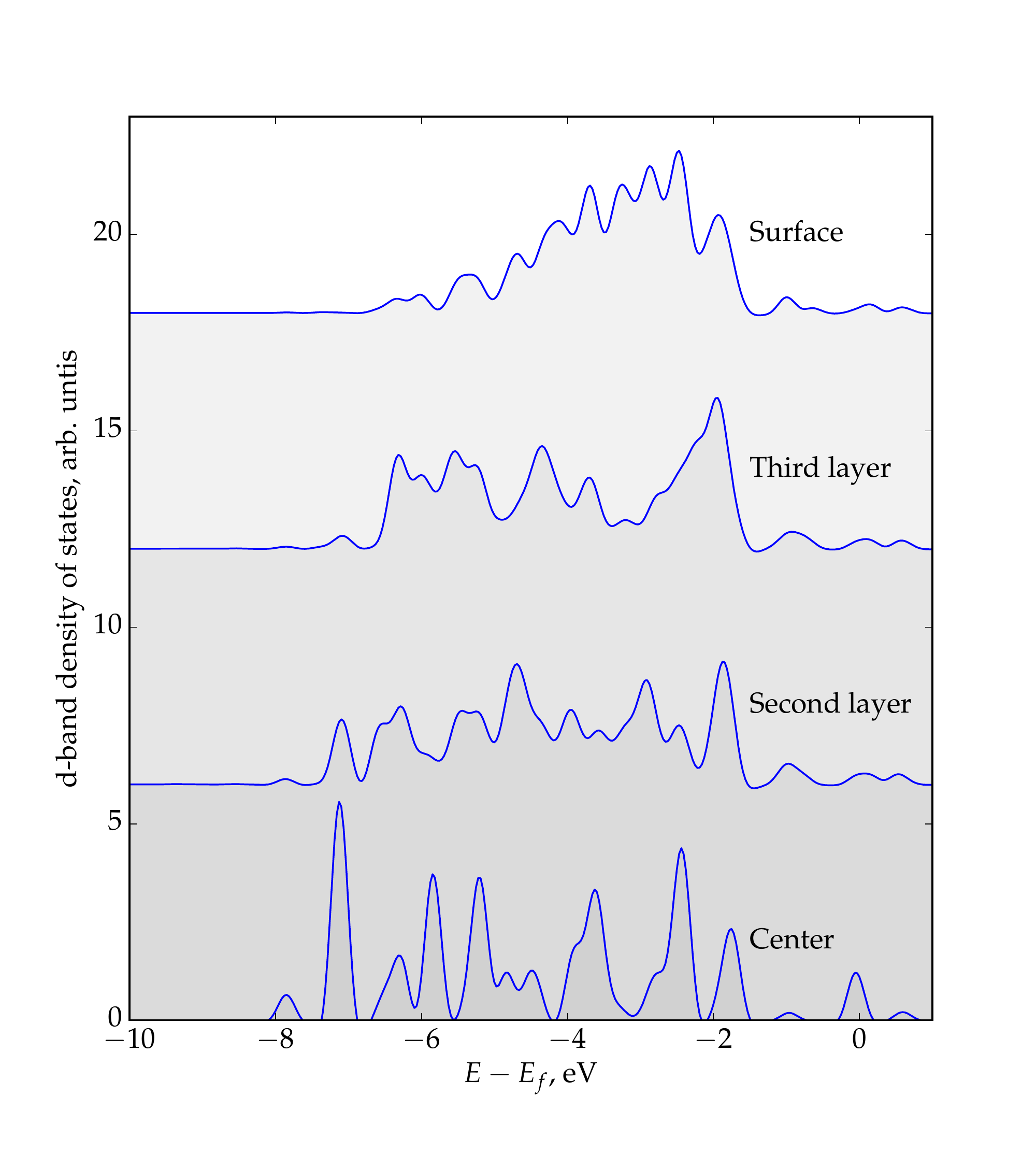}
\label{fig:dos}
\caption{Densities of states for $d$-band of 4 atoms each from one layer (central/first layer, second layer, third layer, surface /last layer) in Au icosahedral nanocluster with 147 atoms. }
\end{figure}

Thus, we conclude that three most significant effects contributing to the 4$f$ shifts in nanoclusters are the number of nearest neighbors or coordination, the distance to the nearest neighbors and the size of the system or confinement effect.
To discuss these effects in more detail for the relatively simple case we show the CLS in a 108 atom cubic nanocluster with bulk lattice parameter in Fig.~\ref{fig:108}. The internal atoms have zero shift as compared to their bulk position while the states for atoms on the surface show big negative CLS.
As it was mentioned early, these effects can be correlated with the behavior of the $d$-band. That allows us to conclude that the CLS is determined by the shift of the electrostatics potential to a very high extent. In Fig.~\ref{fig:dos_conf} the change of the valence band width is clearly observed. The red dots in Fig~\ref{fig:dos_conf} denote the center of the $d$-band.
A decrease in the coordination leads to narrowing of $d$-band as shown in Fig.~\ref{fig:dos} for 147 icosahedral cluster. As it was shown previously \cite{Citrin1983b,Johansson1980}, narrowing of the $d$-band accompanies the negative CLS for metals with more than a half or completely filled band.
The confinement effect also appears together with the band narrowing.

\subsection{\label{sec:Cluster}{Core-level shifts in nanoclusters}}

In order to understand how the structure affects the core-level shifts in nanoclusters we performed CLS calculations in nanoclusters of different size and morphology with magic number of atoms. Magic numbers are the numbers of atoms in perfect cluster structures with all shells filled and all atoms sitting in their ideal position\cite{Harbola1992}.

\paragraph{Icosahedral nanocrystals}

\begin{figure}[ht!]
\includegraphics[width=0.5\textwidth]{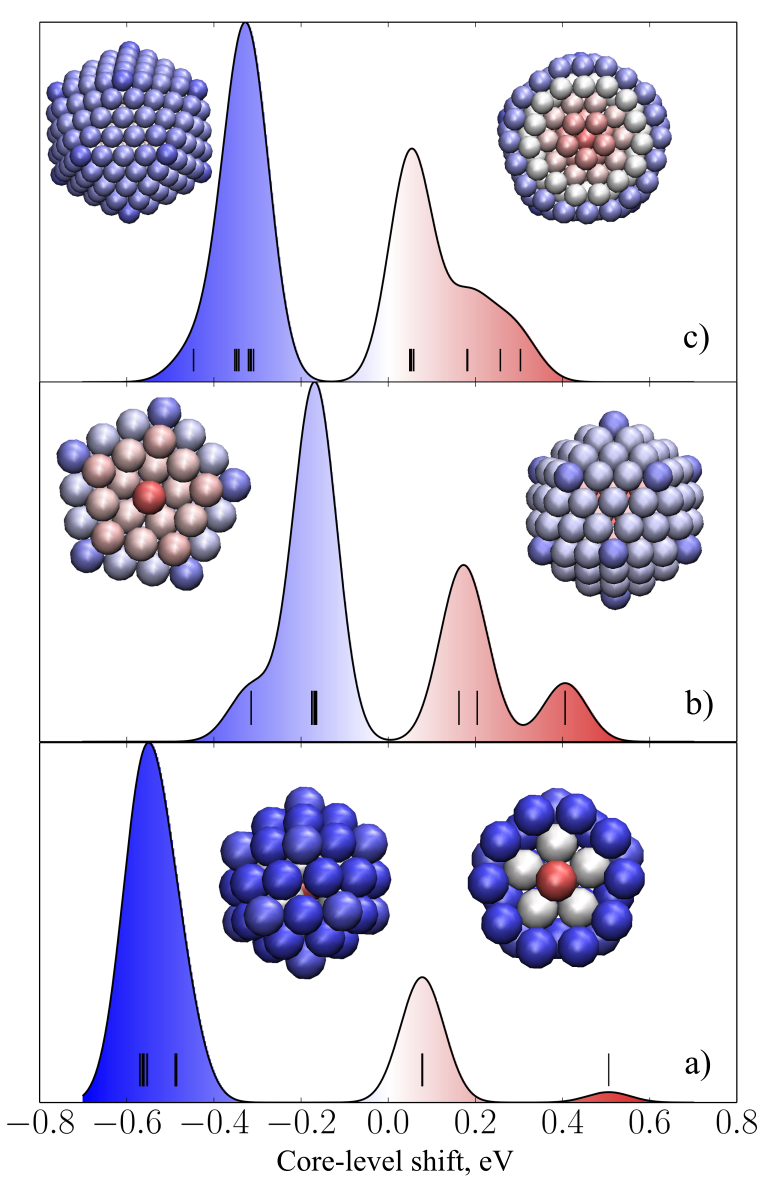}
\label{fig:icos}
\caption{CLS in icosahedral nanoclusters: 55, 147 and 309 atoms. All atoms are colored corresponding to their CLS. Black bars denote calculated CLS. To facilitate comparisons with experimental data, a convolution of the CLS with a 0.05\,eV Gaussian is shown.}
\end{figure}

At small size, it is more energetically favorable to minimize the surface energy of a nanocluster by reducing the surface area. Icosahedral (Ih) structure has the smallest surface area among all cluster morphologies. We have analyzed icosahedral nanoclusters of three sizes: 55, 147 and 309 atoms. In Fig.~\ref{fig:icos}a the icosahedral cluster consisting of 55 atoms is shown. Colors show correspondence of atom position to the CLS. One can see that undercoordinated atoms of the surface have negative CLS of -0.5~eV, while the central atom has a positive CLS of 0.5~eV due to the stresses acting inside the cluster.  The second layer of atoms has negative CLS of -0.1~eV. The next cluster size in Fig. \ref{fig:icos}b consists of 147  atoms or 4 complete layers. The similar trends are observed. Undercoordinated atoms have negative CLS and atoms of the first and the second shell shifted toward higher binding energies. In Fig.~\ref{fig:icos}c the Ih cluster with 309 atoms or 5 layers is shown.  The positive shift peak becomes broader and its contribution increases compared to the surface shifts contribution. The weight of the positive peak should increase as a cube of radius, while the weight of the surface contribution grows as a square. The CLS of the top layer approaches surface CLS of gold. This can be attributed to the fact that confinement effect at such size is negligibly small. In all three clusters the largest shifts have been found on vertex atoms and central atoms. From the series of these three sizes we conclude that the binding energy of core electrons in nanoclusters as a function of the size is non-monotonic and approaches the bulk value with increase of the size.

\paragraph{Decahedral nanocrystals}

\begin{figure}[ht!]
\includegraphics[width=0.5\textwidth]{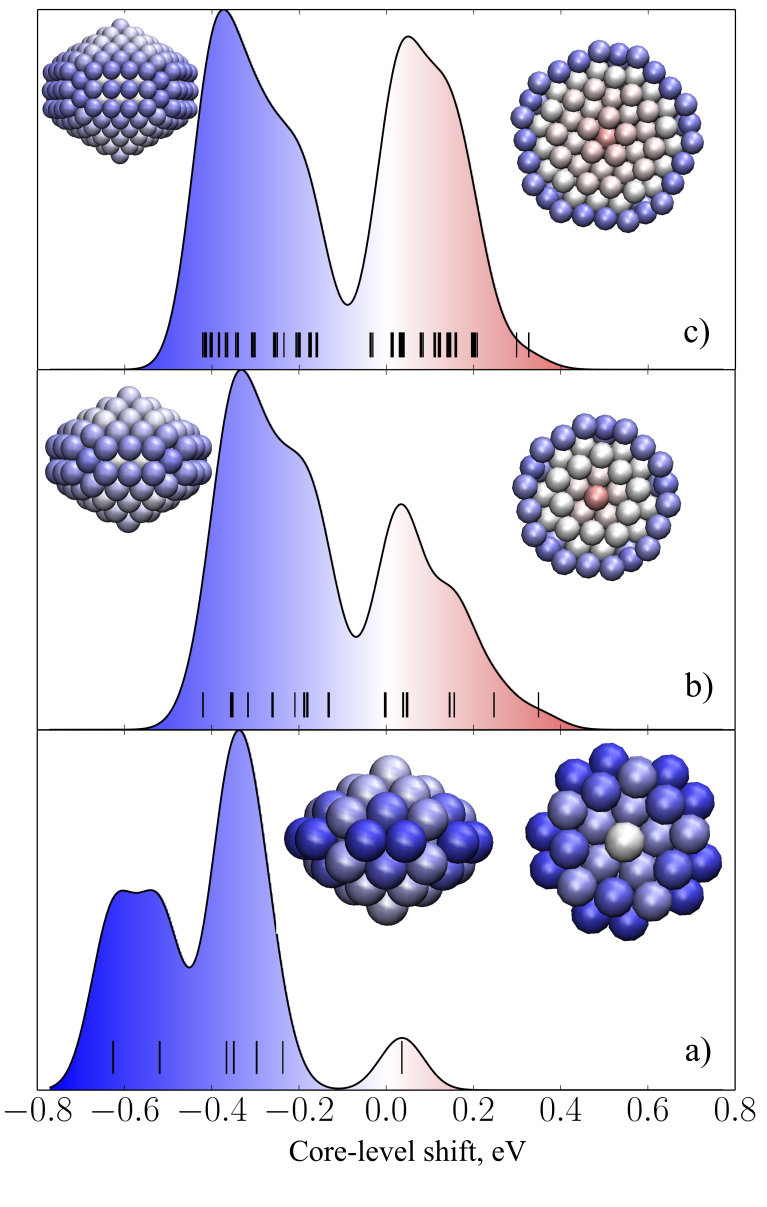}
\label{fig:dec}
\caption{CLS in decahedral nanoclusters: 49, 146 and 318 atoms. All atoms are colored corresponding to their CLS. Black bars denote calculated CLS. To facilitate comparisons with experimental data, a convolution of the CLS with a 0.05 eV Gaussian is shown.}
\end{figure}

In Fig.~\ref{fig:dec} the CLS for three decahedral (Dh) clusters of different size are shown.  The sizes were chosen to be the closest magic number to the size of the icosahedral nanoclusters, which is supposed to make easier the comparison of the nanoclusters with similar number of atoms. Decahedral structure is less symmetrical than icosahedral one, which results in more even distribution of the shifts. A 49 atoms decahedral cluster has many undercoordinated atoms, thus  the weight of the negative CLS is much bigger. With the increase of the size up to 146 atoms the amount of internal atoms grows as well as the weight of the positive shifts. The cluster with 318 atoms has two distinct peaks: one from the internal atoms and one from the surface. Similarly to Ih cluster, surface atoms of Dh at these sizes approach (111) and (100) surface CLS of gold, correspondingly. For all sizes Dh clusters have broader peaks than Ih clusters, which is the result of the symmetry reduction.

\begin{figure}[ht]
\includegraphics[width=0.5\textwidth]{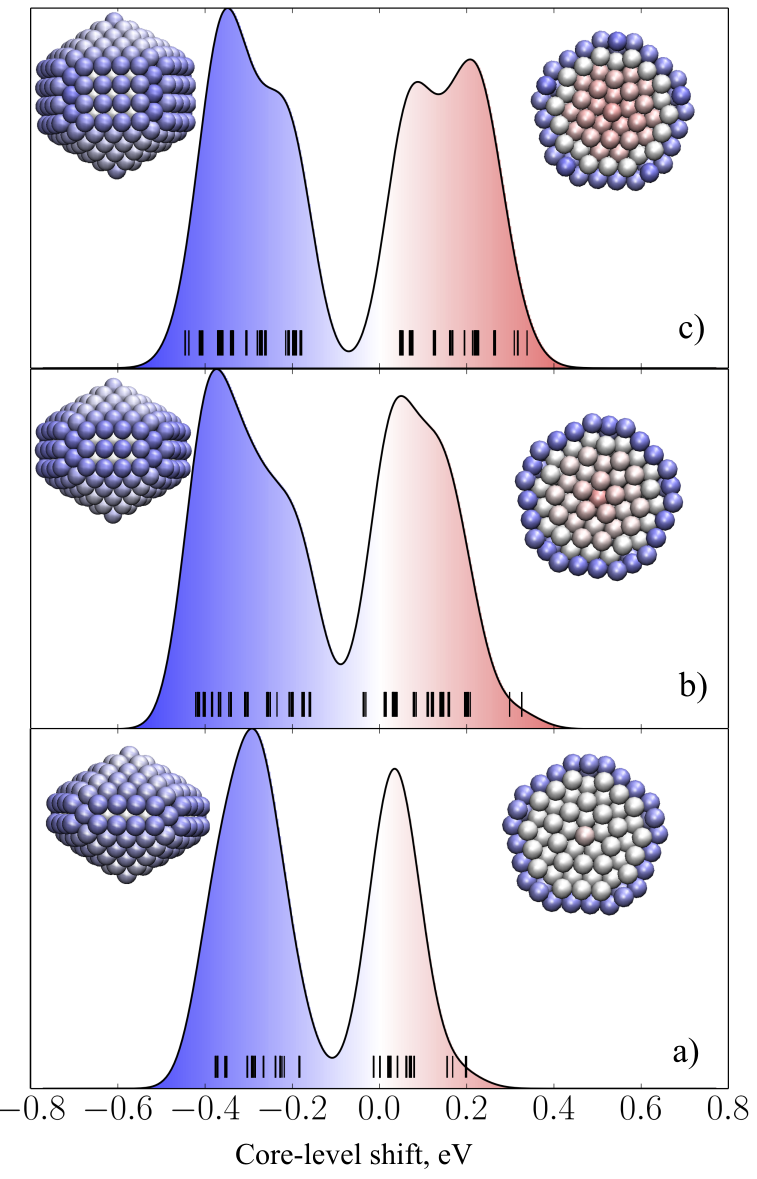}
\label{fig:dec_diff}
\caption{CLS in decahedral nanoclusters: 247, 318 and 389 atoms. All atoms are colored corresponding to their CLS. Black bars denote calculated CLS. To facilitate comparisons with experimental data, a convolution of the CLS with a 0.05 eV Gaussian is shown.}
\end{figure}

While icosahedral geometry can be described by one parameter - number of shells, decahedral structure as less symmetrical structure has more independent parameters. For example, decahedral can be considered as two identical pyramids (top and bottom) with a free number of layers in between them. In order to understand how the structural difference of these configurations will affect the CLS, we calculated the CLS of three Dh cluster with different number of intermediate layers. In fact, the structure of the pyramids is the same for all three structures, while the number of the intermediate layers was changed: 1, 3 and 5. In Fig.~\ref{fig:dec_diff} one can see that the increase of the intermediate layers results in a broadening of the peaks and shift towards higher binding energies. In Dh structures the largest negative CLS are observed on (100) facets and not vertices as in Ih clusters.

\paragraph{Octahedral nanocrystals}

\begin{figure}[ht]
\includegraphics[width=0.5\textwidth]{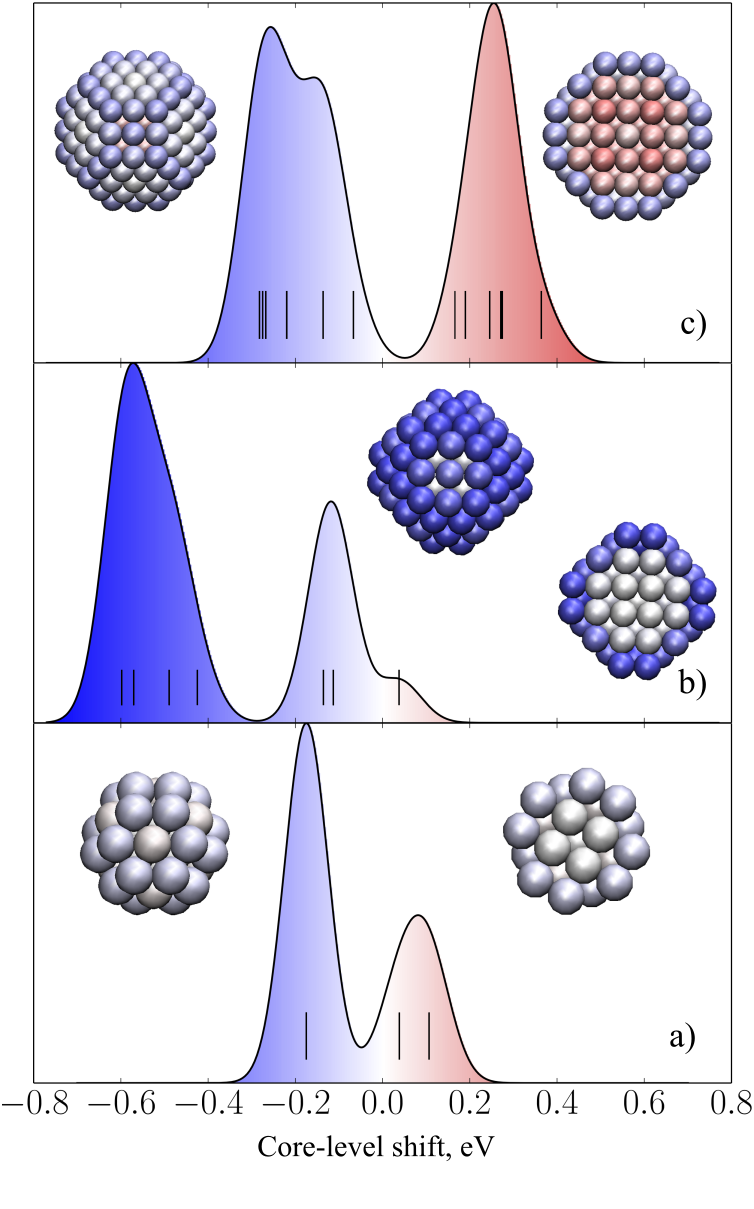}
\label{fig:oct}
\caption{CLS in octahedral nanoclusters: 38, 116 and 201 atoms. All atoms are colored corresponding to their CLS. Black bars denote calculated CLS. To facilitate comparisons with experimental data, a convolution of the CLS with a 0.05 eV Gaussian is shown.}
\end{figure}

Truncated octahedron (TOh) is the least favorable structure at small sizes. The structure of TOh clusters is the most similar to the bulk fcc. Thus, TOh clusters should have least internal stress and the largest surface area. In Fig.~\ref{fig:oct} for all sizes one can see two distinct peaks: surface CLS and internal CLS. At the largest size of 201 atoms the surface peak splits into two: one from $\langle 111 \rangle$ surface with $\sim$-0.2~eV and another from $\langle 100 \rangle$ with $\sim$-0.3~eV.  In 201 atom TOh cluster the lattice parameter is exactly the same as in bulk, but the internal atoms have positive CLS, which is a clear manifestation of the confinement effect.

All clusters have the largest negative CLS on vertices and edges due to coordination effect, while atoms in the center of nanoclusters have large positive CLS due to strains. In most cases, small clusters (around 50 atoms) have the largest surface CLS. In Ih clusters the largest CLS was found for vertex atoms and in Dh and TOh cluster the largest CLS is on (100) facets.  All three considered morphologies have similar signature in the spectra: contributions from surface and bulk are separated into distinct peaks. Dh and TOh clusters have different types of facets, which results in splitting of the surface peak for large clusters, while in Ih clusters all surface atoms have very similar surface CLS.

\section{\label{sec:Conclusion}{Conclusion}}
The evolution of the Au 4$f$ core-state from an individual atom to bulk through cluster has been demonstrated. We have shown how the behavior of the core-level binding energy shift in nanoclusters is governed by three main effects: confinement, stress, and local coordination. In addition, the CLS in the complete screening picture was shown to be very similar to the initial state one. The negative (positive) CLS in the initial state model was explained by the shift of the electrostatic potential.

The difference between XPS features for icosahedral, decahedral and octahedral nanoclusters have been demonstrated. The largest CLS have been found on the edges and vertecies of small nanoclusters.  Understanding of these trends combined with high resolution XPS may allow to distinguish the morphology of the nanoclusters from spectra. Moreover, judging by the magnitude and weights of the surface CLS, TOh with 116 atoms and Dh with 49 atoms could be expected to have the largest absorption energy for other species and to be better for catalysis than other clusters.

\section{\label{sec:Acknowledgments}{Acknowledgments}}

The work was financially supported by the Knut and Alice Wallenberg Foundation through Grants No. 2012.0083 and the Strong Field Physics and New States of Matter 2014-2019 (COTXS). Support by the grant from the Ministry of Education and Science of the Russian Federation (Grant No. 14.Y26.31.0005) is gratefully acknowledged.  W.O. and I.A.A. acknowledge support from the Swedish Government Strategic Research Area in Materials Science on Functional Materials at Link\"{o}ping University (Faculty Grant SFO-Mat-LiU No 2009 00971). The calculations were performed on resources provided by the Swedish National Infrastructure for Computing (SNIC) at the National Supercomputer Centre (NSC) and the supercomputer cluster provided by the Materials Modeling and Development Laboratory at NUST ”MISIS”.

\bibliography{main}{}
\end{document}